# Designer topological insulator with enhanced gap and suppressed bulk conduction in Bi$_2$Se$_3$/Sb$_2$Te$_3$ ultra-short period superlattices

Ido Levy[1,2], Cody Youmans[3,4], Thor Garcia[1,2], Haiming Deng[3,4], Steven Alsheimer[3], Christophe Testelin[5], Lia Krusin-Elbaum[3,4], Pouyan Ghaemi[3,4], Maria Tamargo[1,2,4]

[1] Department of Chemistry, The City College of New York, New York, NY 10031
[2] Chemistry Program, Graduate Center of CUNY, New York, NY 10021
[3] Department of Physics, The City College of New York, New York, NY 10031
[4] Physics Program, Graduate Center of CUNY, New York, NY 10021
[5] Sorbonne Université, CNRS, Institut des NanoSciences de Paris, 4 Place Jussieu, F-75005 Paris, France

Abstract

A novel approach to reduce bulk conductance by the use of short period superlattices (SL) of two alternating topological insulator layers is presented. Evidence for a superlattice gap enhancement (SGE) was obtained from the observed reduction of bulk background doping by more than one order of magnitude, from $1.2 \times 10^{20}$ cm$^{-3}$ to $8.5 \times 10^{18}$ cm$^{-3}$ as the period of Bi$_2$Se$_3$/Sb$_2$Te$_3$ SLs is decreased from 12 nm to 5 nm, respectively. Tight binding calculations show that in the very thin period regime, a significant SGE can be achieved by the appropriate choice of materials. The ultrathin SL of alternating Bi$_2$Se$_3$ and Sb$_2$Te$_3$ layers behaves as a new designer material with a bulk bandgap as much as 60% larger than the bandgap of the constituent layer with the largest bandgap, while retaining topological surface features. Analysis of the weak antilocalization (WAL) cusp evident in the low temperature magneto-conductance of a very thin period SL sample grown confirms that the top and bottom surfaces of the SL structure behave as Dirac surface states. This approach represents a promising and yet to be explored platform for building truly insulating bulk TIs.

There is currently much excitement around the novel physics and potential device applications of 3-D topological insulators (TIs) [1-3]. A bandgap in the bulk, which renders them bulk insulators, along with metallic helical surface states are the principal properties that garner interest in their research. The TI's surface provides a novel electronic state that would harvest exotic quasi-particles, such as Majorana fermions, and other exotic phenomena [4-6]. Groundbreaking potential applications such as quantum computing are envisioned [7]. The most widely studied of these materials is $Bi_2Se_3$ due to its relatively large bandgap and single surface Dirac cone, but other group V-chalcogenides such as $Bi_2Te_3$, $Sb_2Te_3$ are also of great interest [8].

Despite their attractive properties, the still relatively small bandgap of these materials of a few hundred meV, and the ease of formation of electrically active defects[9] result in high bulk conductivities, masking the features of their exotic surface states. In particular, when grown by molecular beam epitaxy (MBE), $Bi_2Se_3$ has n-type conductivity in the bulk[10] and $Sb_2Te_3$ has p-type conductivity [11]. Many attempts to reduce the bulk carrier density in TIs have been previously reported, including modification of the growth conditions and the substrates used [11, 12], impurity compensation doping [13-15], and the growth of mixed alloys [16]. Despite observed reduction of bulk conductivity through these methods, it is evident that significant improvement toward truly insulating bulk TIs need invention of fundamentally new protocols.

Band structure engineering by superlattice (SL) formation has been widely applied to semiconductor physics and devices [17, 18] for many decades. In the field of topological materials, attempts to combine TI layers with trivial semiconductors or combining two topologically trivial materials to get 3D TIs has been explored. Recently, SLs of $Bi_2Se_3/In_2Se_3$ [19-21] have shown that the bandgap and the Dirac cone of the $Bi_2Se_3$ could be modified by the SL structure. Another system, $ZnCdSe/Bi_2Se_3$ SLs, showed that multiple topological surface channels that scale with the number of SL periods could be obtained [22]. A few examples of the growth and properties of heterojunctions and SLs comprised of two TI materials, such as $Bi_2Te_3/Sb_2Te_3$, have also been previously reported [23-25], however the effect of short period TI/TI SL structures on the band structure and carrier density of the resulting materials, or on their topological surface states, has not been appreciably addressed.

In this work we report the growth of short-period SLs of alternating $Bi_2Se_3$ and $Sb_2Te_3$, (TI/TI SLs) which have a type-III or "broken gap" band alignment between the two constituent materials. In type-III band alignment both the valence and the conduction band extrema of one of the constituent materials lie above or below those of the other one, and the bandgaps of the two materials do not overlap[26]. The TI SL structures show a reduction of the bulk carrier density and an increase of the bulk resistivity as a function of the SL period thickness, while still retaining the topological surface states. We interpret this as evidence of new bandgap formation in the SL material that increases with reduced period thickness. Tight binding calculations confirm that for certain materials parameters and TI/TI SL period, new "designer" TI phases can be achieved with bulk bandgaps significantly larger than the bandgaps of either of the constituent layers. This novel approach to materials design and band structure engineering offers much promise for the realization of new TI materials with wider bandgaps than those available in bulk TIs, and represents a promising and yet to be explored novel platform for building truly insulating bulk topological materials.

A series of SL samples consisting of alternating thin layers of $Bi_2Se_3$ and $Sb_2Te_3$ were grown by molecular beam epitaxy (MBE) and their structural properties were investigated by high resolution X-ray diffraction (HRXRD) and transmission electron microscopy (TEM). The SL structures were grown on a $Bi_2Se_3$ buffer layer on (0001) c-plane sapphire substrates with a 0.2° off-cut. A Riber 2300P MBE system was used, equipped with *in-situ* reflection high-energy electron diffraction (RHEED) and background pressures of 3-5x10$^{-10}$ torr during growth. Samples with different SL period thicknesses, as well as different ratios of individual $Bi_2Se_3$ to $Sb_2Te_3$ layer thickness were grown, consisting of seven periods of the two alternating materials. The materials were grown using our previously reported MBE growth procedure that ensures a very smooth and nearly twin free single layer crystals [27]. Details of the growth parameters can be found in the ***Supporting Information Section***.

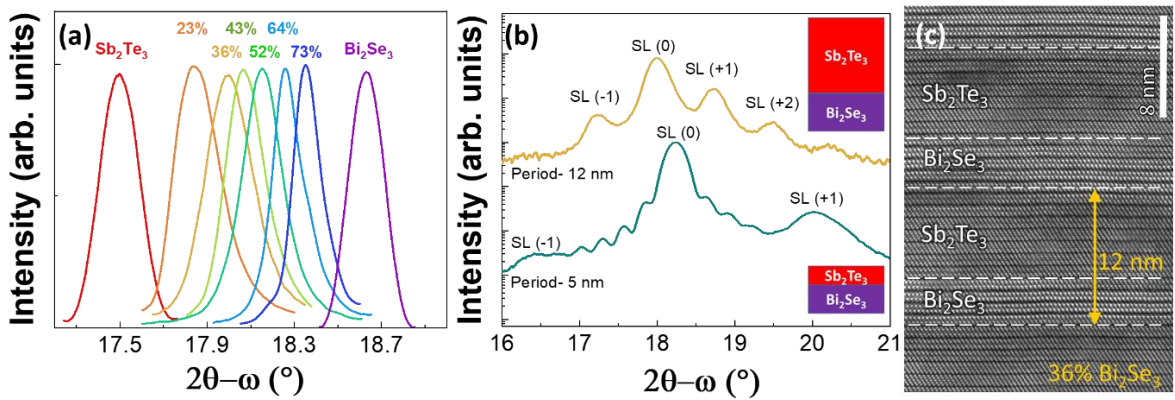

**Figure 1:** a) XRD measurements of the (006) zero-order SL peaks of several of the SL samples studied and of pure $Bi_2Se_3$ and $Sb_2Te_3$ samples. The peak position varied between that of the two pure materials providing a measure of the average composition of the SL. b) Full HR-XRD measurement of the sample with 36% $Bi_2Se_3$ effective composition (top) and 52% effective composition (bottom). From the position of the SL satellite peaks [SL(+1), SL(-1) SL(+2)] the period of the SL could be calculated. **Insets:** period thicknesses with proportional layer thicknesses calculated from XRD. c) HR-TEM cross-section image of the sample with 36% $Bi_2Se_3$ effective composition showing the alternating materials.

Scans of the HRXRD (006) reflection of the zero-order SL peak, SL(0), (Figure 1a and 1b) allowed us to calculate the effective composition of each sample and the thickness of the SL period [28], as well as the individual thicknesses of the $Bi_2Se_3$ and $Sb_2Te_3$ layers. The X-ray diffraction (XRD) measurements were performed using a Bruker D8 Discover diffractometer with a da-Vinci configuration and a Cu K$\alpha_1$(1.5418 Å) source to establish their crystal quality.

The good structural quality was also confirmed by the high resolution (HR) TEM image of the sample with 36% $Bi_2Se_3$ effective composition, which consisted of 4 nm $Bi_2Se_3$ and 8 nm $Sb_2Te_3$, as deduced from the HRXRD data. TEM measurements were performed (EAG Laboratories) using Hitachi HD-2700 Spherical Aberration-Corrected Scanning-TEM with Energy Dispersive X-ray Spectroscopy (EDS). The TEM image (Figure 1c) shows alternating layers with different contrast indicating the two different materials: the lighter contrast corresponding to the $Bi_2Se_3$ layer and the darker one to the $Sb_2Te_3$ layer. The calculated TEM thickness values agree well with the layer thicknesses extracted from the HR-XRD measurements. In addition, the HR-TEM image clearly displays the quintuple layer structure of the individual layers, as well as the high crystallinity of the structure. Although atomically well ordered, the interfaces show some interface steps, or interface roughness, which appears more pronounced on the surfaces of the $Sb_2Te_3$ layers, consistent with observations by ourselves and others, that MBE grown $Sb_2Te_3$ has a higher degree of roughness than $Bi_2Se_3$ [11, 29].

Transport was investigated by Hall Effect measurements, at 10K, using the van-der Pauw[30] (vdP) configuration with Indium contacts on a Lakeshore 7600 electromagnet system. The plot of the Hall resistance ($R_{xy}$) as a function of magnetic field (B) was used to learn the sample conductivity type (n- or p-type). All the samples were measured. Several representative plots are shown in Figure 2a. The corresponding effective composition for each sample is given in the figure. The sign of the slope dictates the conductivity type. We observed a transition of the SL conductivity from n-type to p-type at approximately 42% $Bi_2Se_3$ effective composition.

Carrier density for all the samples was also obtained from the Hall measurements. The data is plotted as a function of the %$Bi_2Se_3$ effective composition in Figure 2b. In this analysis, it was anticipated that the carrier density would reach a minimum at the transition region between n-type to p-type behavior, as charge neutrality was achieved. This transition region is indicated in the figure by the bright region in the background separating the p-type samples from the n-type samples. However, no reduction in the carrier density near the transition region was evident in the data, suggesting that other factors are affecting the variations in conductivity of the samples.

By contrast, a strong correlation is noted in Figure 2c, where the carrier density for the n-type samples is plotted as a function of the SL period. A reduced carrier density for the samples with smaller SL periods is clearly observed. More than an order of magnitude reduction in carrier density was obtained between the samples with larger periods (~12 nm) and those with smaller periods (~5 nm). The low carrier densities of our smaller period TI/TI SLs are comparable to the best values reported for the constituent TI materials grown by MBE directly on sapphire substrates [11, 12, 31, 32]. We believe that these values can be significantly improved, as these results were achieved only by our first efforts. We propose that mastering the properties of TI/TI SLs provides a promising and yet to be explored platform for building truly insulating topological materials.

These observations can be understood based on simple band structure considerations. The expected heterostructure band alignments between $Bi_2Se_3$ and $Sb_2Te_3$ are illustrated in the insets of figure 2c. As previously noted, these two materials have a type-III, or "broken gap" band alignment, [33,34] and the MBE grown $Sb_2Te_3$ is p-type, while the $Bi_2Se_3$ is n-type. [9] As portrayed in the insets, the generation of a SL with small enough periodicity, modifies the band structure and produces new gaps that can be tuned by changing the SL period and thickness[35].

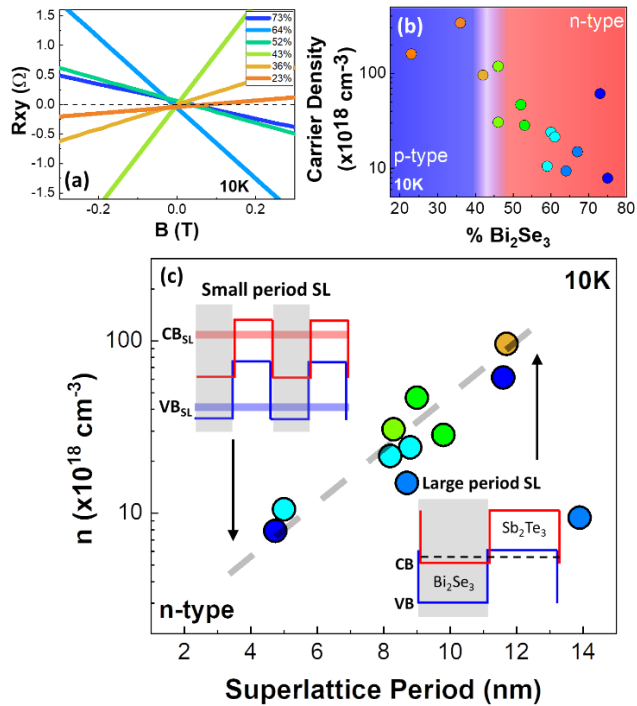

**Figure 2**: a) Hall resistance plots of the SL samples. The effective composition is listed in the legend. The sign of the slope indicates their n-type (blue) or p-type (red) character. b) and c) Carrier concentration of the n-type samples as a function of the %$Bi_2Se_3$ effective composition and SL period thickness, respectively (dashed line is drawn to aid the eye). Insets of figure 2c, Type III band alignment between $Bi_2Se_3$ and $Sb_2Te_3$ and band diagram of a large period SL and a small period SL. The small period SL exhibits the formation of a SL gap, which can be tuned by the period and layer thicknesses. The color of the symbols in figures 2(b) and 2(c) correspond to the colors used in figure 2 (a) for the different samples, which are also the colors assigned to the SL effective compositions in Figure 1.

A better understanding of the properties of these short period TI/TI SLs, and the possibilities of band structure engineering in these novel materials can be obtained from tight binding calculations. Since the alternating layers consist of only a few single layers of each material, we are effectively developing a new type of material and tight-binding is the appropriate method to derive the band structure of our new compound.

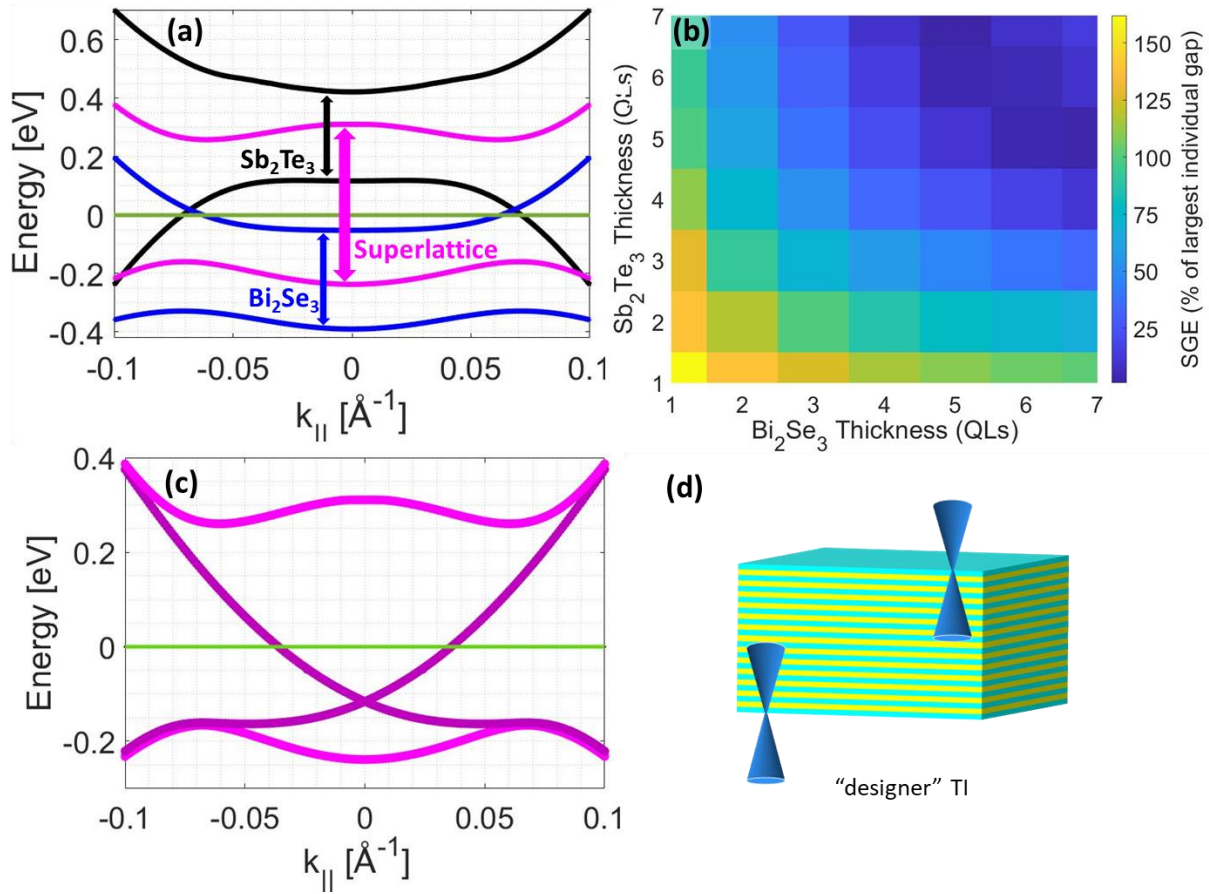

**Figure 3**: (a) Bulk band gaps of $Bi_2Se_3$ (blue), $Sb_2Te_3$ (black), and of an ultrathin binary SL (magenta) of $Bi_2Se_3$ and $Sb_2Te_3$ with respective layer thicknesses $n_1$=1 QL and $n_2$=1 QL. (b) SGE dependence on short-period layer thicknesses. (c) Bulk gap and subgap edge-bands of the SL considered in part (a), (d) Schematic illustrating the resulting "designer TI with enhanced bandgap and top and bottom Dirac surface states.

The low energy properties of both $Bi_2Se_3$ and $Sb_2Te_3$ are well described by the effective k.p Hamiltonian introduced by Zhang et al. [36] Using the parameters from Ref. 37 we built a tight-binding model Hamiltonian. Figure 3(a) shows the resulting highest valence and lowest conduction bands for a SL with $n_1$=1 QL and $n_2$=2 QL, compared with those of pure $Bi_2Se_3$ and $Sb_2Te_3$. The figure highlights an instance of the large superlattice gap enhancement (SGE) expected from the experiments detailed above. A bulk SL bandgap as high as ~160% larger than the $Bi_2Se_3$ bandgap is predicted in Figure 3(b), supporting the proposal of significantly increased bandgaps with decreasing period thickness. It is important to note that the material described by the calculation, consisting of 1QL of $Sb_2Te_3$ and 1 QLs of $Bi_2Se_3$ can no longer be considered as alternating layers of the two bulk materials, as our simple diagrams of figure 2c suggest (as k.p calculations would do, while not pertinent for such thin layers), but rather a new "designer" material. Figure 3(c) demonstrates the preserved topological nature of the composite material by displaying the edge bands at a particular termination of the SL ($Bi_2Se_3$ in our case), within the corresponding bulk band gap. Furthermore, in contrast with the two constituent materials ($Bi_2Se_3$ and $Sb_2Te_3$) the Fermi level in the resulting SL sample lies mid-gap and near the Dirac point, as desired for achieving truly insulating TIs. Lastly, Figure 3(d) presents a diagram of the proposed "designer" TI/TI SL samples with enhanced bulk bandgap and preserved Dirac surface states. Details of the tight binding calculation methodology can be found in the *Supporting Information Section*.

In order to investigate experimentally the presence of the topologically non trivial surface states in these new "designer" TI/TI SLs, as predicted by the tight binding calculations, magneto-conductance (M-C) measurements were performed at 2K on the sample with period thicknesses of 5nm. The M-C measurements were performed in a 14 Tesla Quantum Design Physical property measurement system (PPMS) in 1 mTorr (at low temperature) of He gas. Electrical contacts in the vdP configuration were made with indium bonded on the edge of the thin film. The M-C measurements, shown in Figure 4a, exhibit a weak anti-localization (WAL) cusp, the trademark of 2D transport channels [38,39]. Further analysis of the data was carried out by fitting to the Hikami-Larkin-Nagaoka (HLN) 2D localization theory [40]. From this fit, we extracted a value to the fitting parameter α of 1.0 for the small period (5nm) sample. It has been shown that the parameter α is proportional to the number of 2D channels in the layer, where α= 0.5 represents one channel. Our results suggest that the SL sample behaves like a new TI material with only the top and bottom surfaces of the structure hosting topological surface states, as expected in a bulk 3D TI.

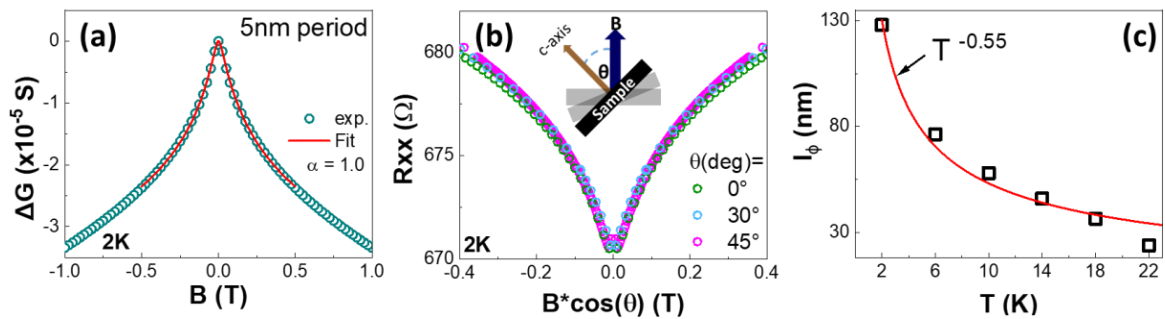

**Figure 4**: (a) Magnetoconductance of samples with different SL period thicknesses. (b) Angle dependent magnetoresistance of the short period SL (sample N) as a function of cosine the field. (c) Temperature dependence of dephasing length $l_\phi$.

Angle dependent magneto-resistance results for the same sample are presented in Figure 4b, where the magneto-resistance, measured with the magnetic field at different angles (θ) normal to the sample surface, is plotted as a function of B(cos θ). All the plots collapse to one curve, suggesting that the 2D conductance channels resulted from the surface electrons. Finally, the magneto-conductance as a function of temperature was also measured. The dephasing length ($l_\phi$) was plotted as a function of temperature and the results are shown in Figure 4c. The data fits well to a temperature dependence of $T^{-0.55}$, very close to the ideal $T^{-0.5}$ dependence predicted by HLN theory for 2D conduction channels [40]. We again conclude that the SL sample with the shorter period behaves as a new "designer" TI phase with a wider bandgap in the bulk and topological surface states at the top and bottom surfaces of the sample. Furthermore, the residual bulk conductivity of the new "designer" TI material can be adjusted by the appropriate choice of SL parameters due to a large SGE possible for ultra-thin period SLs.

This work presents a novel approach to band structure engineering and reduced background doping of topological insulators by the use of ultra-thin period TI/TI SLs. An observed reduction of bulk background doping by more than one order of magnitude as the period of the SLs decreases from 14 nm to 5 nm can be understood on the basis of a SGE possible in type-III "broken gap" heterostructures. This interpretation is supported by tight binding calculations. We show experimentally, from magneto resistance measurements, that the topological surface states of the grown TI/TI SL with the shortest period are in fact preserved. This approach represents a promising and yet to be explored novel platform for building truly insulating bulk topological materials.

**Acknowledgements:**

This work was supported by NSF Grant Nos. DMR-1420634 (MRSEC PAS[3]), HRD-1547830 (IDEALS CREST) and DMR-1824265.

**Author Contributions:**

I.L, T.A.G. and M.C.T. conceived the experiment and executed the crystal growth. I.L. performed HR-XRD and XRR measurements. S.A. and I.L. performed and analyzed the 10K Hall Effect measurements. H.D. and L.K.E performed the 2K magneto conductance measurements. C.Y. and P.G. carried out the analytical theoretical modeling. I.L., C.Y., P.G. and M.C.T. wrote and reviewed the manuscript. All authors contributed to interpretation of the data and discussions.

**Competing Interests:** The authors declare no competing interests

# Supporting Information Section

### A. Growth and Structural Characterization

A series of superlattice (SL) samples consisting of alternating thin layers of $Bi_2Se_3$ and $Sb_2Te_3$ were grown by molecular beam epitaxy. The samples were grown with different SL period thicknesses, as well as different ratios of individual $Bi_2Se_3$ to $Sb_2Te_3$ layer thickness. The SLs were always grown on a $Bi_2Se_3$ buffer layer that was deposited on the sapphire substrate first to ensure a smooth surface for the $Sb_2Te_3$ material growth. All the samples consisted of seven periods of the two alternating materials. The initial $Bi_2Se_3$ buffer layer was grown using our previously reported procedure that ensures a very smooth and nearly twin free single layer crystal. [1]

The samples were characterized with high resolution X-ray diffraction (HR-XRD) to establish their crystal quality. Scans of the (006) reflection of the zero-order SL peak, SL(0) allowed us to determine the effective composition of the samples (%$Bi_2Se_3$). To obtain the effective composition of each SL structure we interpolated the 2θ value of the (006) SL (0) peak between the two end points of pure (006) $Bi_2Se_3$ (at 18.62°) and of pure $Sb_2Te_3$ (at 17.64°) according to equations 1 and 2.

(1) $\quad 2\theta_{SL(0)} = X * 2\theta_{Bi_2Se_3(006)} + (1-X) * 2\theta_{Sb_2Te_3(006)}$,

(2) $\quad \%Bi_2Se_3 = X \times 100 = \frac{SL(0) - Sb_2Te_3}{Bi_2Se_3 - Sb_2Te_3} \times 100$

The full scan of the (006) SL(0) peak for each of the samples show multiple satellite peaks due to their SL structure, indicating that a layered structure with sharp interfaces and a well-defined periodicity was grown. The separation between the satellite peaks allowed us to calculate the thickness of the SL period. [2] Combining the value of the period thickness with the effective composition obtained from the SL(0) peak position, the individual thicknesses of the $Bi_2Se_3$ and $Sb_2Te_3$ layers can be calculated.

### B. Theoretical Modeling

The low energy properties of both $Bi_2Se_3$ and $Sb_2Te_3$ are well described by the effective k.p Hamiltonian introduced by Zhang et al. [3] In a basis of orbital and spin degrees of freedom (represented by Pauli matrices $\tau^i$ and $\sigma^i$, respectively), the corresponding bulk tight-binding Hamiltonian can be written, in terms of the wavevector $p$, as

$$\mathcal{H}_{tb}(\boldsymbol{p}) = C_0 + \sum_i C_i \cos(p_i a_i) + \tau^1 \sum_i v_i \sin(p_i a_i) \sigma^i + \tau^3 \left(M_0 + \sum_i M_i \cos(p_i a_i)\right)$$

where $a_1 = a_2$ is the lattice spacing in the x and y directions within a given layer (4.14 Å for Bi$_2$Se$_3$, and 4.25 Å for Sb$_2$Te$_3$) and $a_3$ is the distance between QLs in the direction perpendicular to the alternating layers (9.55 Å for Bi$_2$Se$_3$, and 10.12 Å for Sb$_2$Te$_3$). The rest of the model fitting parameters for the two materials considered are shown in Table 1. [4]

|         | v$_\parallel$≡v$_1$=v$_2$ | v$_3$ | C$_0$ | C$_\parallel$≡C$_1$=C$_2$ | C$_3$ | M$_0$ | M$_\parallel$≡M$_1$=M$_2$ | M$_3$ |
|---|---|---|---|---|---|---|---|---|
| Bi$_2$Se$_3$ | 0.606 | 0.481 | 3.056 | -1.623 | -0.031 | 7.026 | -3.428 | -0.340 |
| Sb$_2$Te$_3$ | 0.869 | 0.290 | -1.523 | 0.772 | 0.278 | 11.702 | -5.683 | -0.519 |

Table 2: Parameters (given in eV) for the two materials considered in the model.

Rewriting the Hamiltonian in the form $\mathcal{H}_{tb}(\boldsymbol{p}) = H_{os}(\boldsymbol{p}_\parallel) + \Gamma(\boldsymbol{p}_\parallel)e^{-ip_3 a_3} + \Gamma^\dagger(\boldsymbol{p}_\parallel)e^{ip_3 a_3}$, the lattice adaptation along the z-direction within a given SL layer, for a given energy E, is constructed via the recursive relation, $\Gamma\psi_{n-1} + H_{os}\psi_n + \Gamma^\dagger\psi_{n+1} = E\psi_n$, connecting the wave function at the n$_{th}$ lattice site, $\psi_n$, to its nearest neighbors. In between SL layers, the hopping matrices $\Gamma$ are taken to be the average of those within the two layers. The eigen-energies are then obtained through numerical exact diagonalization of the resulting 4N(n$_1$+n$_2$) by 4N(n$_1$+n$_2$) lattice Hamiltonian (where N is number is the total number of supercells, and n$_1$ and n$_2$ are the thicknesses of the two layers forming a supercell).